\documentclass[12pt]{article}

\usepackage[dvips]{graphicx,color}

\begin{document}

\begin{center}

{\Large {\bf Mach's Principle selects 4 space-time dimensions}}

\vspace{2cm}

{Boris L. Altshuler}\footnote[1]{E-mail adresses: baltshuler@yandex.ru \& altshul@lpi.ru}

\vspace{0,5cm}

{\it Theoretical Physics Department, P.N. Lebedev Physical
Institute, \\  53 Leninsky Prospect, Moscow, 119991, Russia}

\vspace{2cm}

{\bf Abstract}

\end{center}
      
Bi-tensor kernel in integral form of Einstein equations realizing Mach's idea 
of non-existence of empty space-times is taken as an inverse of differential 
operator ("Mach operator") defined conventionally as a second variation 
of Einstein's gravity Action over contravariant components of metric tensor. 
The choice of transverse gauge condition used in this definition 
does not influence results of the paper since only transverse and traceless 
tensor modes written on different background space-times 
are studied. Presence of ghosts among modes of Mach operator invalidates 
the integral formulation of Einstein equations. And the demand of absence 
of these ghosts proves to be a selection rule for dimensionality of the background 
space-time. In particular Mach operator written on De Sitter background 
or on the background of so called "Einstein Universe" does not possess 
tensor ghosts only in 4-dimensions. The similar demand gives non-trivial 
formula for dimensionalities of subspaces of the Freund-Rubin background.
                                                             
\vspace{0,5cm}

\newpage
\section{Introduction. Integral form of Einstein \\ equations and definition of Mach operator}

\qquad There are no dynamical answer yet to the questions: why observed Universe is 3+1 dimensional? Why, if we suppose higher dimensions of equal rights at the Big Bang, only 3 space dimensions expand to large volume? Beginning from the Ehrenfest pioneer work \cite{Ehrenfest} a number of important observations about the privileged character of 3+1 space-time were made. In frames of string theory and branes' dynamics interesting attempts to explain the expansion of 3 dimensions based upon the observations that $2+2<5$ and $4+4<10$ were made in \cite{Vafa} and \cite{Karch} correspondingly. Also promising recent publication says that this explanation is known to supecomputer \cite{Monte}. In the present paper the special role of (3+1) is revealed in totally different context.

In \cite{Altsh1}-\cite{Altsh2} the integral representation of solutions of Einstein equations

\begin{equation}
\label{1}
R_{AB}-\frac{1}{2}g_{AB}R=\kappa T_{AB}
\end{equation}
was proposed:

\begin{equation}
\label{2}
g_{AB}(x)=\kappa N \int G^{\it{ret}}_{AB}{}^{PQ}(x,y|g)T_{PQ}(y) \sqrt{-g}\, d^{N}y.
\end{equation}
which we write down here for space-time of arbitrary dimensionality $N$ ($A,B=0,1,2 \cdots\, (N-1)$; signature: $-\,+\,+\,+\cdots$; $\kappa$ is gravitational constant in $N$ dimensions).
The integral form (\ref{2}) is the vivid formulation of Mach's conjecture of the relativity of non-inertial movements \cite{Mach} in formulation given by Einstein \cite{Einstein} of space-time being totally created by matter. (\ref{2}) is evidently a selection rule excluding in particular empty solutions of Einstein equations (\ref{1}) for which (\ref{2}) symbolically comes to impossible relation $1=0$. Also it is easy to show that (\ref{2}) excludes asymptotically flat space-times.

Kernel $G^{\it{ret}}_{AB}{}^{PQ}(x,y|g)$ in (\ref{2}) (indices {A,B} refer to the point $x$; $P,Q$ - to $y$) is a bi-tensor Green function on the background space-time described by the same metric given by (\ref{2}) and is the retarded solution of the equation

\begin{equation}
\label{3}
E_{AB}^{CD}G_{CD}{}^{PQ}(x,y)=\delta_{A}^{P}\delta_{B}^{Q}\frac{\delta^{N}(x-y)}{\sqrt{-g}},
\end{equation}
where covariant differential operator $E_{AB}^{CD}$ is defined on the background (\ref{2}). This operator must satisfy the simple condition:

\begin{equation}
\label{4}
E_{AB}^{CD}g_{CD}=N(R_{AB}-\frac{1}{2}g_{AB}R),
\end{equation}
then its action upon (\ref{2}) with account of (\ref{3}) immediately gives (\ref{1}). 

There are many doubts and questions as regards to integral formulation (\ref{2}) of Einstein equations. One of questions was put to the author by John Archibald Wheeler (in 1968 at the Second International Gravitational Conference in Tbilisi) who said: "Why don't you include energy-momentum of gravitational waves in the source in the RHS of (\ref{2})?". I came back now to this old stuff not because I found out an answer, but because for the naturally defined (see (\ref{5}) below) differential operator $E_{AB}^{CD}$ in (\ref{3}) the demand of validity of integral representation of solutions of Einstein equations unexpectedly proves to be a selection rule for the dimensionality of space-time.

We define the differential operator $E_{AB}^{CD}$ taking the second variation of the Einstein Action $\int R\sqrt{-g}$:

\begin{eqnarray}
\label{5}
&&E_{AB,CD}h^{CD}=\frac{2}{\sqrt{-g}}\,\frac{\delta^{2}(R\sqrt{-g})}{\delta g^{AB}\delta g^{CD}}h^{CD}= \nonumber \\
&&\frac{2}{\sqrt{-g}}\,\frac{\delta[\sqrt{-g}(R_{AB}-\frac{1}{2}\,g_{AB}R)]}{\delta g^{CD}}\,h^{CD}=  \nonumber \\  
&&\left[-g_{AC}g_{BD}\nabla^{2}-2R_{AC,BD}+R_{AC}g_{BD}+R_{BC}g_{AD}+ \frac{1}{2}g_{AB}g_{CD}\nabla^{2}+ \right.  \nonumber  \\
&&\left. R_{AB}g_{CD}+g_{AB}R_{CD}-g_{AC}g_{BD}R-\frac{1}{2}g_{AB}g_{CD}R \right]h^{CD},
\end{eqnarray}
where $\nabla^{2}=g^{PQ}\nabla_{P}\nabla_{Q}$ is D'Alambertian, and $h_{CD}$ are small variations of metric

\begin{equation}
\label{6}
g_{CD} \to g_{CD}+h_{CD}
\end{equation}
subject to the transverse gauge condition

\begin{equation}
\label{7}
\nabla_{B}(h_{A}^{B}-\frac{1}{2}\delta_{A}^{B}h_{C}^{C})=0.
\end{equation}
It is worthwhile to note immediately that in this paper we, following \cite{Gibb}, \cite{KI1}, \cite{KI2}, basically consider gauge-invariant tensor variations of the stationary background metrics of class (\ref{13}) - see below. And since tensor modes are transverse and traceless by their definition the conclusions of this paper do not depend on the choice of gauge condition (\ref{7}).

First four terms of operator $E_{AB,CD}$ in square brackets in the RHS of (\ref{5}) are the standard Lichnerowicz operator $\triangle_{L}$ which gives the variation of Ricci tensor in the gauge (\ref{7}): $2\delta R_{AB}=(\triangle_{L}h)_{AB}=-\nabla^{2} h_{AB}-2R_A{}^C{}_B{}^{D}h_{CD}+R_{A}^{C}h_{BC}+R_{B}^{C}h_{AC}$. Substitution of $g_{CD}$ instead of $h_{CD}$ in (\ref{5}) gives (\ref{4}); thus Green function defined by (\ref{3}), (\ref{5}) may be used in formulation of Mach's Principle in a form (\ref{2}). We call differential operator (\ref{5}) {\it {Mach operator}}.

Boundary conditions imposed by integral form (\ref{2}) upon solutions of Einstein equations (\ref{1}) are easily received if we express energy-momentum tensor in the RHS of (\ref{2}) from (\ref{1}), (\ref{4}): $N \kappa T_{PQ}=E_{PQ}^{CD}g_{CD}$, and then integrate (\ref{2}) by parts. This (with account $\nabla_{M}g_{CD}\equiv 0$) gives

\begin{equation}
\label{8}
\oint \left[\sqrt{-g(y)}\nabla_{N_{y}}G^{\it{ret}}_{AB}{}^{Q}{}_{Q}(x,y)\right]\,dS^{N_{y}}=0
\end{equation}
which is the integral over boundary of space-time. Fulfillment of (\ref{8}) guarantees the validity of (\ref{2}).

Now we come to the formulae which will be used in the bulk of this paper.

Essentially more strong conditions than (\ref{8}) are imposed upon metric satisfying integral representation (\ref{2}) if we demand that "neighboring" solutions of Einstein equations also are purely inhomogeneous. Thus we present integral form for the small variations of metric $\delta g_{AB}=h^{(e)}_{AB}$ (symbol $(e)$ means that this is a solution of linear variation of Einstein equations (\ref{1}) on the background metric $g_{AB}$ satisfying (\ref{2})). Variation of (\ref{2}) gives (symbolically) $\delta g = \delta G \cdot T + G \cdot \delta T$. The first term is calculated from variation of (\ref{3}) preserving the retarded nature of the Green function: $\delta G= - G\cdot \delta E \cdot G$, thus with account of (\ref{2}): $\delta G \cdot T = -G \cdot \delta E \cdot g$. Second term is received from variation of (\ref{1}), (\ref{4}): $G \cdot \delta T = G \cdot \delta E \cdot g + G\cdot E \cdot \delta g$. This chain of variations gives finally:

\begin{equation}
\label{9}
h^{(e)}_{AB}(x) = \int G^{\it{ret}}_{AB}{}^{CD}(x,y|g)E_{CD}^{PQ}h^{(e)}_{PQ}(y)\sqrt{-g}\,d^{N}y.
\end{equation}
Here, according to the definition of $E_{AB}^{CD}$ in (\ref{5}):

\begin{equation}
\label{10}
E_{AB}^{CD}h^{(e)}_{CD}=\frac{2}{\sqrt{-g}}\delta(\sqrt{-g}\kappa T_{AB}), 
\end{equation}
which is just the variation of Einstein equations (\ref{1}).

Eq. (\ref{9}) comes to identity if we act upon it with differential operator $E_{AB}^{CD}$. "Machian" absence of the "free term" in the RHS of (\ref{9}) means the fulfillment, in analogy with (\ref{8}), of the following boundary condition:

\begin{eqnarray}
\label{11}
\oint\{\sqrt{-g(y)}[(\nabla_{N_{y}}G^{\it{ret}}_{AB}{}^{PQ}(x,y))\,(h^{(e)}_{PQ}-\frac{1}{2}g_{PQ}h^{(e)K}_{K}) - \nonumber \\
G^{\it{ret}}_{AB}{}^{PQ}(x,y)\nabla_{N_{y}}(h^{(e)}_{PQ}-\frac{1}{2}g_{PQ}h^{(e)K}_{K})]\}\,dS^{N_{y}}=0.   
\end{eqnarray}

Demand of validity of (\ref{9}), or equivalently of (\ref{11}), for any Einstein $h^{(e)}_{AB}$ variation of background metric $g_{AB}$ is a strong selection rule for this background metric. Formulae (\ref{9}), (\ref{11}) (with account of (\ref{3}), (\ref{5}), (\ref{10})) are explored below in a number of simple models.

\section{Ghosts of Mach operator invalidate the integral representation: Einstein Universe \\ as an example of preferred 4 dimensions}

Existence of the ghost solutions of variation (\ref{10}) of the Einstein equations would mean as usual the instability of the background space-time $g_{AB}$. Whereas existence of ghosts of Mach operator (\ref{5}), i.e. presence of ghosts among solutions of the homogeneous equations

\begin{equation}
\label{12}
E_{AB}^{CD}u_{CD}=0,
\end{equation}
results in non-fulfillment of the "retarded" boundary condition (\ref{11}) imposed by the integral representation (\ref{9}). We shall demonstrate it below on a simplest example of the background Einstein Universe where Mach operator $E_{AB}^{CD}$ proves to be "non-ghost" for tensor modes only in 4 dimensions. 

But first let us make several notes about general approach of this paper. We shall follow analyses of \cite{Gibb}, \cite{KI1}, \cite{KI2} and study a number of stationary background space-time models described by the metric of type 

\begin{equation}
\label{13}
ds^{2}=g_{ab}dx^{a}dx^{b}+r^{2}(x)d\sigma _{n}^{2};
\end{equation}
here $x^{a}$ are coordinates of the $m$-dimensional space-time, $a=0,1\ldots (m-1)$; $d\sigma_{n}^{2}=\gamma_{ij}dx^{i}dx^{j}$ is the metric of the $n$-dimensional $G_{n}$-invariant space with normalized constant sectional curvature $K=0, \pm 1$. So the dimension of the whole space-time is $N=m+n$.

Again following \cite{Gibb}, \cite{KI1}, \cite{KI2} we consider the simplest case of the gauge-invariant transverse and traceless tensor modes of Eq-s (\ref{12}) (and (\ref{10})) written on the backgrounds (\ref{13}):

\begin{equation}
\label{14}
u_{ab}=u_{ai}=0; \,\, u_{ij}=\tilde{u}_{ij},  \,\, \nabla_{j}\tilde{u}_{i}^{j}=0, \, \, \tilde{u}_{i}^{i}=0
\end{equation}
(the same for $h^{(e)}_{AB}=\tilde {h}^{(e)}_{AB}$ in (\ref{10})). For the backgrounds (\ref{13}) tensor modes separate from vector and scalar modes in Eq-s (\ref{12}) or (\ref{10}). As it is shown in \cite{KI1}, \cite{KI2} the analyses of the gauge-invariant vector and especially scalar modes is a complicated task, while for tensor modes the task is quite simple; detailed description of tensor excitations of different backgrounds of type (\ref{13}) is given in \cite{Gibb}.

Tensor modes (\ref{14}) "live" on the $n$-dimensional subspace of (\ref{13}). In Section 4 we shall also consider tensor modes "living" only on $m$-dimensional subspace of (\ref{13}): $u_{ij}=u_{ai}=0$, $u_{ab}=\tilde{u}_{ab}$ which are transverse and traceless in $m$-dimensional space-time.

It will be shown that demand of absence of ghosts among solutions of (\ref{12}) is a strong selection rule for background space-times and in particular essentially restricts their dimensionality.

Now we come to the simplest example of the Einstein Universe background.
In 1917 Einstein introduced $\Lambda$-term in his gravity equations and built with it a model of space-time $T\times S^{3}$ known as the Einstein Universe. Let us consider metric of $N=(1+n)$ Einstein Universe $T\times S^{n}$ 

\begin{equation}
\label{15}
ds^{2}= - dt^{2}+r_{0}^{2}d\Omega_{n}^{2}.
\end{equation}
This is metric (\ref{13}) for $x^{a}=t$, $r = r_{0} = const$ and $K= +1$. 

And let us consider metric (\ref{15}) as a background in Eq-s (\ref{12}) and (\ref{10}) written for tensor modes (\ref{14}) of $u_{AB}$ and correspondingly of $h^{(e)}_{AB}$ "living" on sphere $S^{n}$:
\begin{equation}
\label{16}
\tilde{u}_{i}^{j}=f(t)v_{i}^{j}(x^{k}), \,\,\, \tilde{h}^{(e)j}_{i}=f^{(e)}(t)v_{i}^{j}(x^{k}),
\end{equation}
$v_{i}^{j}(x^{k})$ are tensor eigenmodes of the Laplace-Beltrami operator on $n$-sphere of unit radius: $\triangle_{S^{n}}v_{i}^{j}= l(l+n-1)-2; \,\, l=1,2\ldots; \,\, n=3,4\ldots$ (tensor modes do not exist on 2-sphere).

Then Eq-s (\ref{10}), (\ref{12}), with account of (\ref{5}) written on the background (\ref{15}), come correspondingly to the following simple equations for scalar functions $f^{(e)}(t)$ and $f(t)$ introduced in (\ref{16}) (we omit here momentum index $l$):

\begin{equation}
\label{17}
\left[\frac{d^{2}}{dt^{2}}+\mu^{(e)2}\right] f^{(e)}(t)=0, \,\,\, \mu^{(e)2}r_{0}^{2}=l(l+n-1);
\end{equation}

\begin{equation}
\label{18}
\left[\frac{d^{2}}{dt^{2}}+\mu^{2}\right] f(t)=0, \,\,\, \mu^{2}r_{0}^{2}=l(l+n-1) - 2 - n(n-3),
\end{equation}
here $l\ge 1$ and $n\ge 3$.

Since $\mu^{(e)2}$ in (\ref{17}) is positive the Einstein Universe (\ref{15}) is stable. At the same time $\mu^{2}$ in (\ref{18}) may be negative. For the most "ghosts-threatening" lowest value of momentum number ($l=1$) we have from (\ref{18}):

\begin{equation}
\label{19}
\mu^{2}r_{0}^{2}=4n-2-n^{2}
\end{equation}
which is positive for $n=3$ and negative for $n\ge 4$. Thus Mach operator (\ref{5}) 'selects' (1+3)-dimensional Einstein Universe.

Let us show that existence of ghosts solutions of (\ref{18}) invalidates integral form (\ref{9}). With account of (\ref{16}) in this simple case (\ref{9}) looks as:

\begin{eqnarray}
\label{20}
&&f^{(e)}(t)=\int D^{\it{ret}}(t-t')\left[\frac{d^{2}}{dt'^{2}}+\mu^{2}\right] f^{(e)}(t')\, dt' \nonumber \\
&&= \int D^{\it{ret}}(t-t')\left[\mu^{2}-\mu^{(e)2}\right] f^{(e)}(t')\, dt',
\end{eqnarray}
where the last equality is written from comparison of (\ref{17}) and (\ref{18}). And for the corresponding boundary (here - initial) condition (\ref{11}) we receive:

\begin{equation}
\label{21}
\lim_{t'\to -\infty}\left[\frac{d D^{\it{ret}}(t-t')}{d t'}f^{(e)}(t')- D^{\it{ret}}(t-t')\frac{df^{(e)}(t')}{dt'}\right] = 0.
\end{equation}

Green function $D^{\it{ret}}(t-t')$ in (\ref{20}), (\ref{21}) obeys the equation (cf. (\ref{3}), (\ref{12}), (\ref{18})):

\begin{equation}
\label{22}
\left[\frac{d^{2}}{dt^{2}}+\mu^{2}\right] D^{\it{ret}}(t-t')=\delta (t-t'),
\end{equation}
and is easily found explicitly:

\begin{equation}
\label{23}
D^{\it{ret}}(t-t')=\frac{1}{2i\mu}\theta(t-t')\left[e^{i\mu(t-t')-e^{-i\mu(t-t')}}\right].
\end{equation}

For $\mu^{2}>0$ integral form (\ref{20}) may be Fourier transformed in a standard way and becomes identity. Whereas for $\mu^{2}<0$ one of two solutions of (\ref{18}) determining behavior of Green function (\ref{23}) exponentially increases at $t' \to - \infty$. Thus initial condition (\ref{21}) (hence integral form (\ref{20})), can not be fulfilled when there are ghosts among solutions of Eq. (\ref{18}).

With this extremely simple model we demonstrated the connection of ghosts of Mach operator (\ref{5}) written on stationary background with non-fulfillment of "machian" integral form (\ref{9}) for Einstein variations of this background. In what follows, having this in mind, we just explore the "dimensionality dependence" of the appearance of ghosts of Mach operator (\ref{5}) on some typical background space-times.

\section{$dS_{N}$ background: $N=4$ selected}

For this highly symmetric background ($R_{AB}=c(N-1)g_{AB}$) Eq-s (\ref{12}), where $E_{AB}^{CD}$ is given in (\ref{5}), separate for traceless (${\tilde u}_{A}^{B}$) and scalar (${\bar u}=u_{D}^{D}$) components of $u_{A}^{B}$:

\begin{equation}
\label{24}
(\triangle _{L}{\tilde u})_{A}^{B} - cN(N-1){\tilde u}_{A}^{B}=(-\nabla^{2} -cN(N-3)){\tilde u}_{A}^{B}=0,
\end{equation}

\begin{equation}
\label{25}
(\nabla^{2} - cN(N-1)){\bar u}=0.
\end{equation}
($\triangle _{L}$ - Lichnerowicz operator, $\nabla^{2}$ - D'Alambertian in $N$ dimensions).

From (\ref{24}) it is seen that on the De Sitter background ($c>0$) mass squared of traceless modes of Mach operator becomes negative for space-time dimension $N\ge 4$. Actually, as we'll show now, the ghost-problems of Mach operator at this background begin for $N\ge 5$.

The only difference of Eq. (\ref{24}) from the analogous equation (\ref{10}) for tensor variations of Einstein equations is in the term $cN(N-1)$ in the first part of (\ref{24}). For the "mass shell" variation (\ref{10}) of De Sitter space-time this term must be changed by $2c(N-1)$ (cf. Eq. (24) in \cite{Gibb}); hence in \cite{Gibb} mass squared of traceless modes is non-negative and de Sitter space is stable as expected. 

Let us now look at the ghosts of Mach operator (\ref{5}) written on the De Sitter background which metric we take here in a form (\ref{13}), where $x^{a}= t, r$:

\begin{equation}
\label{26}
ds^{2}=-(1-c\,r^{2})dt^{2}+\frac{dr^{2}}{1-c\,r^{2}}+r^{2}d\Omega_{n}^{2},
\end{equation}
($d\Omega_{n}^{2}$ is metric of round sphere $S^{n}$).

We consider tensor modes on a sphere:

\begin{equation}
\label{27}
u_{a}^{j}=u_{a}^{b}=0, \,\, u_{i}^{j}=\varphi (r)\,e^{iEt} v_{i}^{j}, \,\, u_{i}^{i}=0, \,\, \nabla_{j}u_{i}^{j}=0,
\end{equation}
for definition of $v_{i}^{j}$ and its spectrum - see (\ref{16}) and comments there; and we again omit everywhere the momentum index $l$.

By changing variable $r$ to Regge-Wheeler type dimensionless coordinate $y$ and rescaling field $\varphi$:

\begin{equation}
\label{28}
dy=\frac{\sqrt{c}dr}{1-cr^{2}}, \,\,\,\,\,\, r=\frac{1}{\sqrt{c}}\tanh y, \,\, \,\, \varphi = r^{-n/2}\Phi
\end{equation}
($0<y<\infty$) (\ref{24}) comes to a Schrodinger-type equation

\begin{equation}
\label{29}
-\frac{d^{2}\Phi}{dy^{2}}+\left[\frac{4l(l+n-1)+n^{2}-2n}{4\sinh^{2}y}-\frac{\beta(n)}{4\cosh^{2}y}\right]\Phi = \frac{1}{c}E^{2}\Phi.
\end{equation}

Coefficient $\beta(n)$ in potential in square brackets is function of dimensionality $n$ and Eq. (\ref{29}) actually embraces four different cases of interest with different dependences $\beta (n)$ - two for tensor modes and two for vector modes (definition of vector modes see in \cite{Gibb}-\cite{KI2}): 

\begin{eqnarray}
\label{30}
&&\beta_{T}=5n^{2}+6n; \,\, \beta_{T}^{(e)}=n^{2}+2n; \nonumber 
\\
&&\beta_{V}=5n^{2}+2n; \,\, \beta_{V}^{(e)}=n^{2}-2n.
\end{eqnarray}
Here $\beta_{T}(n)$ is received from Mach operator (\ref{5}) and $\beta_{T}^{(e)}(n)$ - from variation (\ref{10}) of Einstein equations; both (\ref{5}) and (\ref{10}) being written on the background (\ref{26}) (for $\beta_{T}^{(e)}$ - cf. Formulae (26), (27) of \cite{Gibb}). And similar expressions are put down in (\ref{30}) for gauge-invariant vector modes $\beta_{V}$ (for Mach operator) and $\beta_{V}^{(e)}$ (for variation of Einstein equations - cf. Formula (5.15) in \cite{KI1}).

Potential $V(y)= [\cdots]$ in square brackets in (\ref{29}) is, as expected, non-negative for $\beta=\beta^{(e)}_{T,V}$, i.e. for tensor and vector variations of de-Sitter background. But it is not the case for corresponding eigenmodes of Mach operator. 

Undesirable ghost exists if for $E^{2}<0$ in (\ref{29}) the normalization condition is valid:

\begin{equation}
\label{31}
\int \varphi^{2}\frac{r^{n}}{1-cr^{2}}\,dr = \int_{0}^{\infty}\Phi^{2}dy < \infty.
\end{equation}

Ghost appears if negative potential well of $V(y)=[\cdot]$ in (\ref{29}) is sufficiently deep. Fortunately there is exact solution of (\ref{29}) which clarifies the word "sufficiently":

\begin{equation}
\label{32}
\Phi = (\tanh y)^{l+n/2}(\cosh y)^{-\gamma}, \,\,\, \gamma=\frac{1}{2}\sqrt{1+\beta}-\left(l+\frac{n}{2}+\frac{1}{2}\right),
\end{equation}
with the ghost-like negative energy squared $E^{2}=-c \gamma^{2} < 0$. 

This solution meets normalization condition (\ref{31}) if $\gamma > 0$ in (\ref{32}). This is evidently not the case for "Einstein" values of $\beta=\beta^{(e)}_{T,V}$ in (\ref{30}), hence there are no normalized ghost modes among Einstein variations of de Sitter metric. For tensor modes of Mach operator $\beta=\beta_{T}$ in (\ref{30}) normalization condition $\gamma>0$ looks as:

\begin{equation}
\label{33}
\sqrt{5n^{2}+6n+1}>2l+n+1 \,\, (l\ge 1,\,\, n\ge 3).
\end{equation}
In particular for $n=3$ (i.e. for the 5 dimensional space-time (\ref{26})) (\ref{33}) comes to $2-l > 0$, hence normalization condition (\ref{31}) is fulfilled for the ghost-like tensor mode with $l=1$. 

Thus there are ghosts among tensor modes of Mach operator (\ref{5}) written on the 5-dimensional (and higher than 5 dimensions) De Sitter backgrounds. That is integral form (\ref{9}) written for tensor modes on De-Sitter background of five and more dimensions is plagued by the ghosts of Green function $G$.

For vector modes, $\beta=\beta_{V}$ in (\ref{30}), and for the minimal value of momentum number $l=2$ (for $l=1$ vector modes are Killing vectors of $S^{n}$ which time-dependence in two dimensions $(r, \, t)$ may be gauged away \cite{KI1}, \cite{KI2}) normalization condition $\gamma>0$ for the ghost-like solution (\ref{32}) looks as $(n^{2}-2n-6)>0$ which is fulfilled for $n\ge 4$. Thus Mach operator on De Sitter background is plagued by vector ghosts in 6 and more dimensions.

At the same time integral form (\ref{2}) refers only to scalar part of Mach operator (\ref{5}). Thus Mach's boundary condition (\ref{8}) may be written in this case, with account of (\ref{25}), in a simple scalar form

\begin{equation}
\label{34}
\oint \left[\sqrt{-g(y)}\nabla_{N_{y}}G(x,y)\right]\,dS^{N_{y}}=0,
\end{equation}
where scalar Green function obeys $(\nabla^{2} - cN(N-1)) G(x,y)=\frac{\delta^{(N)}(x-y)}{\sqrt{-g}}$ ((cf. (\ref{25})). It is not difficult to show that (\ref{34}) is valid for de Sitter background metric (\ref{26})  for any space-time dimension $N$.

\section{$AdS_{m}\times S^{n}$ background}

\qquad Let us look first at the pure $AdS$. In this case $c<0$ in (\ref{24}), (\ref{25}) and there are no ghost problems in (\ref{24}) since mass term of traceless modes is non-negative for $N\ge3$. However scalar modes $\bar u$ of Mach operator are ghosts at this background since their negative mass squared in (\ref{25}) is below the Breitenlohner-Freedman bound \cite{Breiten}, \cite{Breiten2}:

\begin{equation}
\label{35}
\frac{m^{2}}{|c|}= - N(N-1) < - \frac{(N-1)^{2}}{4}
\end{equation}
for all $N$.

It is possible also to show that in the Randall-Sundram model \cite{RS}
presence of the $Z_{2}$-symmetric co-dimension one brane results in the ghost bound state of tensor mode of Mach operator in the $\delta$-function negative well potential of the brane. 

More interesting is to consider the properties of Mach operator (\ref{5}) when in (\ref{12}) the background is the Freund-Rubin $AdS_{m}\times S^{n}$ space-time \cite{Freund}:

\begin{equation}
\label{36}
ds^{2}=dz^{2}+ e^{-2Hz}\eta_{\mu\nu}dx^{\mu}dx^{\nu}+r_{0}^{2}d\Omega_{n}^{2},
\end{equation}
which stability was investigated in \cite{Wolfe} (where instead of $S^{n}$ more general compact space $M_{n}$ was considered). Metric (\ref{36}) is a special form of metric (\ref{13}); $\eta_{\mu\nu}$ is Minkowski metric in $(m-1)$ dimensions, $d\Omega_{n}^{2}$ is metric of unit sphere. Components of Ricci tensor of space-time (\ref{36}) are ($a,\, b$, enumerate coordinates $z,\, x^{\mu}$ of $AdS_{m}$ and $i,\, j$ enumerate coordinates of $S^{n}$):

\begin{equation}
\label{37}
R_{ab}= - (m-1)H^{2}g_{ab}, \,\,\, R_{ij}= \frac{n-1}{r_{0}^{2}}g_{ij}=\frac{(m-1)^{2}}{n-1}H^{2}g_{ij},
\end{equation}
and total scalar curvature is given by:

\begin{equation}
\label{38}
R=\frac{m-1}{n-1}(m-n)H^{2}.
\end{equation}

Homogeneous Eqs. (\ref{12}) (with differential operator $E$ from (\ref{5}) on the background (\ref{36})) written for spherical tensor modes $u_{i}^{j}=\varphi_{(n)}(z, x^{\mu})v_{i}^{j}(x^{k})$ on $S^{n}$ (cf. (\ref{14}), (\ref{16})) and spherical scalar modes $u_{\mu}^{\nu}=\varphi_{(m)}(z, x^{\mu})v_{\mu}^{\nu}(x^{k})$ on $AdS_{m}$ (we again omit everywhere the spherical momentum index $l$) come to two equations for scalars $\varphi_{(n)}$ and $\varphi_{(m)}$ correspondingly:

\begin{equation}
\label{39}
(\triangle(m) - M_{(n),(m)}^{2})\varphi_{(n),(m)}=0,
\end{equation}
where $\triangle(m)=g^{ab}\nabla_{a}\nabla_{b}$ is D'Alambertian on $AdS_{m}$ and effective masses $M_{(n)}$ and $M_{(m)}$ are:

\begin{eqnarray}
\label{40}
&&M_{(n)}^{2}=\frac{l(l+n-1)-2 +2n}{r_{0}^{2}}- R=   \nonumber  \\
&&\left[\frac{(m-1)^{2}}{(n-1)^{2}}(l(l+n-1)-2+2n)- \frac{m-1}{n-1}(m-n)\right] H^{2},
\end{eqnarray}
here $l=1,2\ldots$; \,\, $n\ge3$;

\begin{eqnarray}
\label{41}
&&M_{(m)}^{2}=\frac{l(l+n-1)}{r_{0}^{2}}- 2(m-1)H^{2}-R =     \nonumber \\
&&\left[\frac{(m-1)^{2}}{(n-1)^{2}}\,l(l+n-1) -2(m-1) - \frac{m-1}{n-1}(m-n)\right] H^{2},
\end{eqnarray}
here $l=0,1\ldots$; \,\, $n\ge 2$. In (\ref{39}), (\ref{40}) scalar curvature $R$ is taken from (\ref{38}).

It is easily seen that Breitenlohner-Freedman condition for $AdS_{m}$ \cite{Breiten}, \cite{Breiten2} $M^{2}\ge [- (m-1)^{2} / 4] H^{2}$ which guarantees absence of the ghosts-solutions of Eq. (\ref{39}) is always fulfilled for $M_{(n)}^{2}$ (\ref{40}). Whereas for $M_{(m)}^{2}$ (\ref{41}) it gives for lower spherical mode $l=0$:

\begin{equation}
\label{42}
mn+9 \ge 5(m+n).
\end{equation}
If this condition is violated Mach differential operator (\ref{5}) gives ghost solutions of Eq. (\ref{12}) written on the $AdS_{m}\times S^{n}$ background (\ref{36}) for tensor modes $u_{\mu}^{\nu}$ of the $AdS_{m}$ subspace of $AdS_{m}\times S^{n}$.

Minimal dimension of space-time $AdS_{m}$ permitted by (\ref{42}) is $m=6$, in this case (\ref{42}) is fulfilled for $n\ge 21$, i.e. for total dimension $N=m+n \ge 27$.

For $m=n$, i.e. for the $AdS_{n}\times S^{n}$ background, (\ref{42}) gives $n\ge 9$. In this case $R=0$ (see (\ref{38})) and for tensor modes under consideration Mach operator (\ref{5}) comes to the Lichnerowicz operator $\triangle _{L}$ given by the first four terms in the RHS of (\ref{5}). Thus $n\ge 9$ (i.e. $N=2n \ge 18$) is a condition of absence of ghosts of tensor modes of Lichnerowicz operator written on the $AdS_{n}\times S^{n}$ background.

\section{Discussion}

\qquad Main results of this paper are given by Formulae (\ref{19}) 
and (\ref{33}) which show that (3+1)-dimensional "Einstein universe" 
and De Sitter space-time are singled out by the demand of absence of 
ghosts of Mach operator (\ref{5}). However this is just a mathematical 
observation. To connect result (\ref{33}) for De Sitter universe with 
the possible dynamical answer to the 
nagging question "Why only 3 space dimensions expand during inflation?" 
is an open problem.

Yes, results of Sec.2 demonstrate that ghosts of Mach operator 
invalidate integral form (\ref{9}) written in particular on the De Sitter 
background of higher than 4 dimensions. But in this paper, 
as well as in all preceding papers \cite{Altsh1} - \cite{Altsh2}, the integral 
formulation of Einstein equations was written down "by hand" without 
dynamical grounds for it. To find these grounds is perhaps a way to find 
an answer to the "nagging question" above.

The ideas of "gravity without gravity" (rephrasing Weeler's favourite 
saying) or of "space-time totally created by matter" (which comes up to 
Mach's idea of relativity of accelerated movements) look quite dynamical. 
And hystorically Mach's ideas inspired Einstein for creation of General 
relativity, which however did not exclude empty ("non-machian") solutions 
of Einstein equations. 

In string theory graviton is a dynamical excitation of more fundamental 
object and background space-time is Bose condensate of these excitations, 
and Einstein's gravity Action comes up as an effective one.
However string theory suffers from plethora of admissible backgrounds which 
deprives it of physical predictability. String theory evidently needs 
additional selection rules. Can integral forms (\ref{2}), (\ref{9}) 
be among such rules?

Also nice condition (\ref{42}) of absence of ghosts of Mach operator 
taken on the Freund-Rubin $AdS_{m}\times S^{n}$ background is so far 
just a numerology of dimensionalities which may become science in case 
dynamical grounds for integral formulation of Einstein equations will 
be found out.

\section*{Acknowledgements} Author is grateful for fruitful discussions to 
participants of the Quantum Field Theory Seminar in the Theoretical Physics 
Department, Lebedev Physical Institute.

\end{document}